\documentclass[aps,twocolumn,groupedaddress]{revtex4-1}

\usepackage{graphicx}
\usepackage{bm}
\usepackage{tabularx}
\usepackage{amsmath}

\begin{document}

\preprint{}

\title{Antiferromagnetism, superconductivity, and pseudogap in three-layered high-$T_c$ cuprates Ba$_2$Ca$_2$Cu$_3$O$_6$(F,O)$_2$ probed by Cu-NMR   
}

\author{Sunao Shimizu}
\email[]{e-mail: shimizu@nmr.mp.es.osaka-u.ac.jp}
\author{Shin-ichiro Tabata}
\author{Hidekazu Mukuda}
\author{Yoshio Kitaoka}
\affiliation{Department of Materials Engineering Science, Osaka University, Osaka 560-8531, Japan }
\author{Parasharam M. Shirage}
\author{Hijiri Kito}
\author{Akira Iyo}
\affiliation{National Institute of Advanced Industrial Science and Technology (AIST), Umezono, Tsukuba 305-8568, Japan}

\date{\today}

\begin{abstract}
We report on the phase diagram of antiferromagnetism (AFM) and superconductivity (SC) in three-layered Ba$_2$Ca$_2$Cu$_3$O$_6$(F,O)$_2$ by means of Cu-NMR measurements. It is demonstrated that AFM and SC uniformly coexist in three-layered compounds as well as in four- and five-layered ones. The critical hole density $p_c$ for the long range AFM order is determined as $p_c\simeq$ 0.075, which is larger than $p_c\simeq$ 0.02 and 0.055 in single- and bi-layered compounds, and smaller than $p_c\simeq$ 0.08-0.09 and 0.10-0.11 in four- and five-layered compounds, respectively. 
This variation of $p_c$ is attributed to the magnetic interlayer coupling which becomes stronger as the stacking number of CuO$_2$ layers increases; that is, the uniform coexistence of AFM and SC is a universal phenomenon in underdoped regions when a magnetic interlayer coupling is strong enough to stabilize an AFM ordering.
In addition, we highlight an unusual pseudogap behavior in three-layered compounds -- the gap behavior in low-energy magnetic excitations collapses in an underdoped region where the ground state is the AFM-SC mixed phase.
\end{abstract}


\maketitle

\section{Introduction}Since the discovery of the superconductivity (SC) in hole-doped copper-oxides, cuprates have shown the highest SC transition temperature $T_c$ of many other superconducting materials. Especially, it is known that three-layered compounds have the highest $T_c$ in cuprates \cite{Scott,IyoTc}. For example, Hg-based three-layered HgBa$_2$Ca$_2$Cu$_3$O$_{8+\delta}$ (Hg1223) exhibits $T_c$=133 K at an ambient pressure \cite{Schilling} and reaches $T_c$=164 K at a pressure of 31~GPa \cite{Gao}. To understand electronic properties of three-layered compounds would be a key to addressing the mechanism of high-$T_c$ SC, but few systematic investigations with tuning hole density $p$ have been reported.

We have reported the phase diagram of multilayered cuprates with four and five CuO$_2$ planes~\cite{MukudaJPSJ,ShimizuJPSJ}, where SC uniformly coexists with a long-range AFM order in underdoped regions~\cite{ShimizuFIN}. 
The phase diagrams for multilayered cuprates are different from well-established ones for single-layered La$_{2-x}$Sr$_x$CuO$_4$ (LSCO) \cite{Keimer,JulienLSCO,JulienReview} and bi-layered YBa$_2$Cu$_3$O$_{6+\delta}$ (YBCO) \cite{Sanna,Coneri}; the long-range AFM order in LSCO (YBCO) is strongly suppressed with the hole density $p$ of $\sim$ 0.02 (0.055), while it survives up to $p$ $\sim$ 0.08-0.09 (0.10-0.11) in four-layered (five-layered) cuprates.
The fact that a phase diagram depends on the stacking number $n$ of CuO$_2$ layers is probably due to a magnetic interlayer coupling being stronger with increasing $n$.
Note that in multilayered compounds, the high-$T_c$ SC takes place above 100 K \cite{IyoTc} despite such strong AFM correlations.   
In order to investigate an interplay between AFM and SC, it is important to unravel the phase diagram of AFM and SC in three-layered compounds, which hold the highest $T_c$ in cuprates. 

In this paper, we report systematic Cu-NMR studies on three-layered high-$T_c$ cuprates Ba$_2$Ca$_2$Cu$_3$O$_6$(F,O)$_2$ (0223F) in a wide hole-density region. The $p$ values of the samples are estimated from the Knight-shift measurement as listed in Table \ref{t:ggg}.
The $^{63}$Cu-NMR spectra and the nuclear spin-lattice relaxation rate show that in 0223F 
an AFM order exists up to $p$ $\sim$ 0.075 and uniformly coexists with SC. We also report that the spin-gap behavior in low-energy magnetic excitations disappears in the paramagnetic normal state in the AFM-SC mixed region.  

\section{Experimental details}

Polycrystalline powder samples used in this study were prepared by high-pressure synthesis, as described elsewhere \cite{Iyo1,Iyo2}. The three samples of Ba$_2$Ca$_2$Cu$_3$O$_6$(F$_y$O$_{1-y}$)$_2$ (0223F) exhibit a systematic change in $T_c$, as listed in Table \ref{t:ggg}, as the nominal amount of fluorine F$^{1-}$ (i.e., $y$) increases. The heterovalent substitution of F$^{1-}$ for O$^{2-}$ at the apical sites (see Fig.~\ref{fig:NMR}(a)) decreases $p$ in apical-F compounds \cite{Iyo2,ShimizuJPSJ,Shirage}. Note that it is difficult to investigate the actual fraction of F$^{1-}$ and O$^{2-}$ \cite{Shirage,ShimizuPRL,ShimizuPRB}. Here, $T_c$ was uniquely determined by the onset of SC diamagnetism using a dc SQUID magnetometer, and powder X-ray diffraction analysis shows that these compounds comprise almost a single phase. For NMR measurements, the powder samples were aligned along the $c$-axis in an external field $H_{ex}$ of $\sim$ 16 T and fixed using stycast 1266 epoxy. The NMR experiments were performed by a conventional spin-echo method in the temperature ($T$) range of 1.5 $-$ 300 K with $H_{ex}$ perpendicular to the $c$-axis ($H_{ex}\perp c$).
 
\section{Results}

\subsection{$^{63}$Cu-Knight shift}
\begin{figure}[htpb]
\begin{center}
\includegraphics[width=1.0\linewidth]{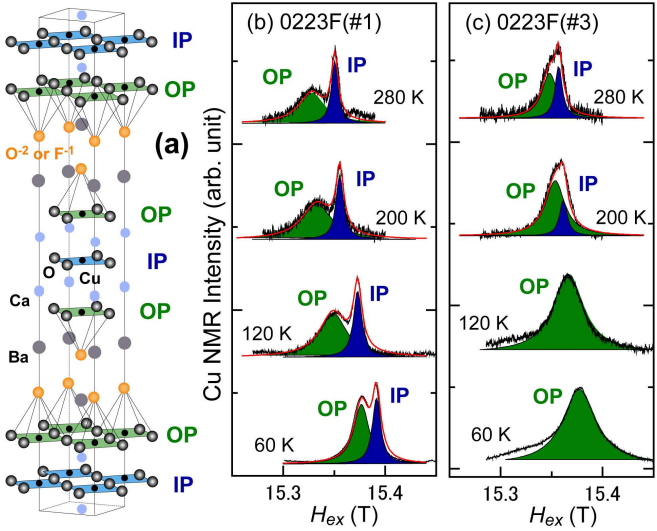}
\end{center}
\caption{\footnotesize (color online)  (a) Crystal structure of three-layered Ba$_2$Ca$_2$Cu$_3$O$_6$(F,O)$_2$. The heterovalent substitution of F$^{1-}$ for O$^{2-}$ at apical sites decreases the hole density $p$. The outer and inner CuO$_2$ planes are denoted as OP and IP, respectively. (b,c) $^{63}$Cu-NMR spectra for 0223F($\sharp$1) and 0223F($\sharp$3) with $H_{ex}$ $\perp$ $c$. In 0223F($\sharp$3), the spectrum of IP is lost at low temperatures due to the development of AFM correlations.}
\label{fig:NMR}
\end{figure}

\begin{table}[htbp]
\caption[]{List of Ba$_2$Ca$_2$Cu$_3$O$_6$(F$_y$O$_{1-y}$)$_2$ used in this study. $T_c$ was determined by the onset of SC diamagnetism using a dc SQUID magnetometer. Hole densities for each CuO$_2$ plane, $p$(OP) and $p$(IP), are evaluated from the Cu Knight shift measurement (see text).}
\label{samples}
\begin{center}
{\tabcolsep = 3.5mm
\renewcommand\arraystretch{1.3}
  \begin{tabular}{ccccc}
    \hline\hline
Sample           &   $y$ &      $T_c$(K) &    $p$(OP)  &   $p$(IP)       \\
    \hline
0223F($\sharp$1) &   0.6 &   120  &      0.156  &    0.111        \\
0223F($\sharp$2) &   0.8 &    102 &     0.113   &    0.087        \\
0223F($\sharp$3) &   0.9 &     81 &      0.086  &    0.073
        \\

    \hline\hline
    \end{tabular}}
 \end{center}
\label{t:ggg}
\end{table}
Figures \ref{fig:NMR}(b) and \ref{fig:NMR}(c) indicate typical $^{63}$Cu-NMR spectra of the central transition (1/2 $\Leftrightarrow$ $-$1/2) for 0223F($\sharp$1) and for 0223F($\sharp$3), respectively. The field-swept NMR spectra were measured at $H_{ex}\perp c$, and the NMR frequency $\omega_0$ was fixed at 174.2 MHz. As shown in Fig.~\ref{fig:NMR}(a), 0223F has two kinds of CuO$_2$ planes: outer CuO$_2$ plane (OP) and inner plane (IP). The two peaks in the NMR spectra correspond to OP and IP. The assignment of NMR spectra to OP and IP was already reported in previous literature \cite{Julien,ZhengTl2223,Kotegawa2001}. The Cu-NMR spectra of OP and IP are detected in 0223F($\sharp$1), whereas the spectrum of IP is lost in 0223F($\sharp$3) at low temperatures, as shown in Fig.~\ref{fig:NMR}(c). This is because AFM correlations develop towards an AFM order upon cooling, as in the case of four- \cite{ShimizuJPSJ} and five-layered compounds~\cite{MukudaJPSJ}. 
\begin{figure}[tpb]
\begin{center}
\includegraphics[width=0.85\linewidth]{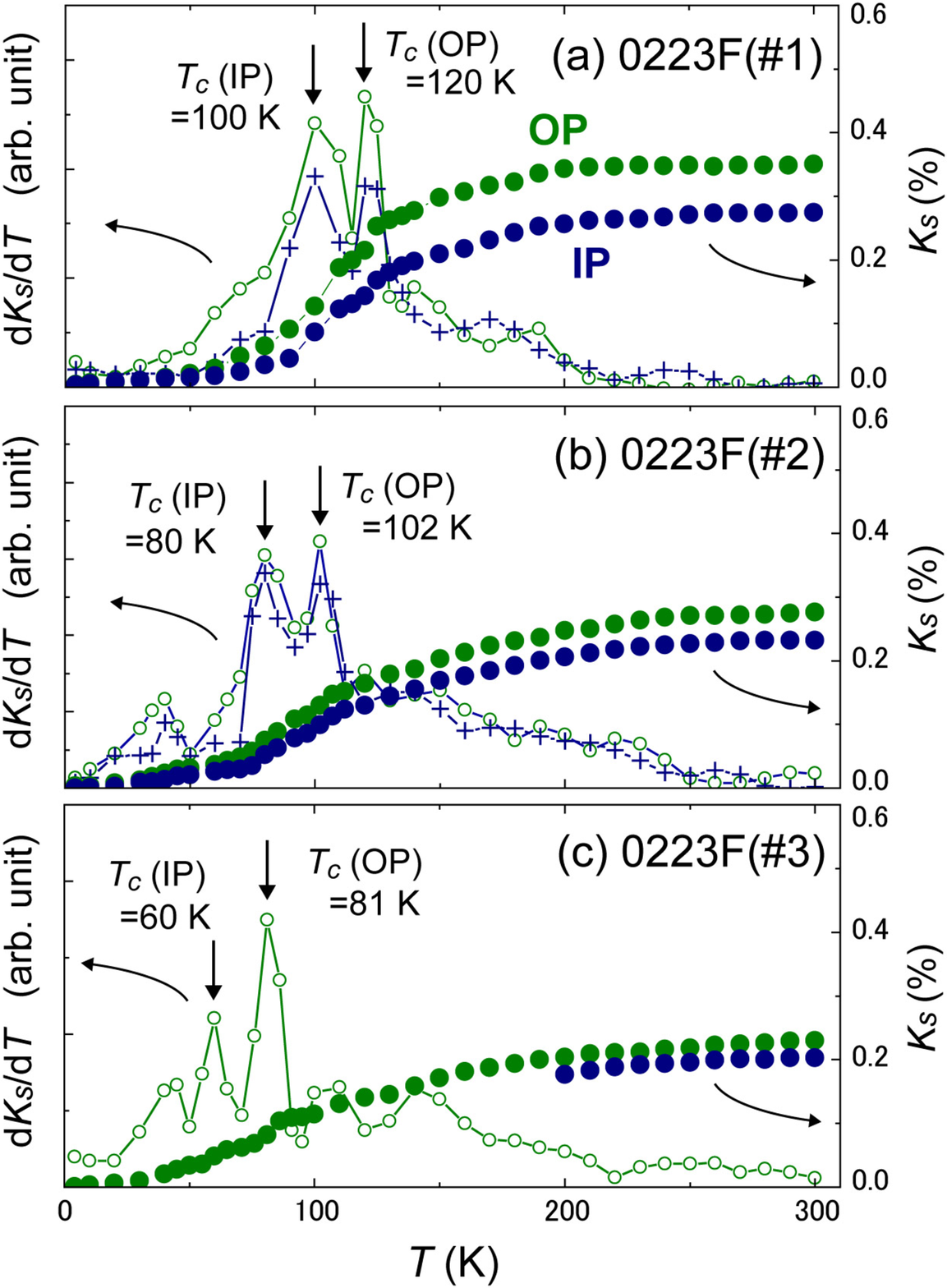}
\end{center}
\caption{\footnotesize (color online)  $T$-dependences of $^{63}$Cu Knight shift $K_s(T)$ with $H_{ex} \perp c$ and of $dK_s(T)/dT$ for (a) 0223F($\sharp$1) (cited from Ref.~\cite{ShimizuP}), (b) 0223F($\sharp$2), and (c) 0223F($\sharp$3). The data of $dK_s(T)/dT$ for OP and IP are marked as open circles ($\circ$) and crosses ($+$), respectively.}
\label{fig:KS}
\end{figure}

According to the second order perturbation theory for the nuclear Hamiltonian with $H_{ex}\perp c$ \cite{Abragam,TakigawaNQRshift}, the NMR shifts of the spectra in Figs.~\ref{fig:NMR}(b) and \ref{fig:NMR}(c) consist of the Knight shift $K$ and the second order quadrupole shift. The NMR shift is expressed as 
\begin{equation}
\frac{\omega_0 - \gamma_N H_{res}}{\gamma_N H_{res}} = K+\frac{3\nu_Q^2}{16(1+K)}\frac{1}{(\gamma_N H_{res})^2}~, 
\label{eq:shift}
\end{equation}
where $\gamma_N$ is a nuclear gyromagnetic ratio, $H_{res}$ the NMR resonance field, and $\nu_Q$ the nuclear quadrupole frequency. In order to subtract $K$ from the total NMR shift, we estimated $\nu_Q$ from NQR (nuclear quadrupole resonance) spectra measured at $T$=1.5 K and $H_{ex}$=0 T, which are shown in Fig.~\ref{fig:NQR}.

In high-$T_c$ cuprates, the quantity $K$ comprises a $T$-dependent spin part $K_s(T)$ and a $T$-independent orbital part $K_{orb}$ as follows:
\begin{equation}
K= K_{s}(T)+K_{orb}. 
\label{eq:K}
\end{equation}
The $T$-dependences of $K_s(T)$ with $H_{ex}\perp c$ for 0223F($\sharp$1), 0223F($\sharp$2), and 0223F($\sharp$3) are displayed in Figs.~\ref{fig:KS}(a), \ref{fig:KS}(b), and \ref{fig:KS}(c), respectively. Here, $K_{orb}$ was determined as $\sim$ 0.22 \%, assuming $K_{s}(T) \simeq 0$ at a $T=$ 0 limit. 
As shown in the figures, the value of $K_s(T)$ at room temperature decreases as $p$ decreases from 0223F($\sharp$1) to 0223F($\sharp$3). 
The values of $p$ at OP and at IP, which are summarized in Table \ref{t:ggg}, are separately evaluated by using the relationship between $K_s$(300 K) and $p$ \cite{ShimizuP}.
In Fig.~\ref{fig:KS}, $K_s(T)$ decreases upon cooling down to $T_{c}$ for all samples due to the reduction of the density of states at the Fermi level in association with the opening of pseudogaps \cite{Yasuoka,REbook}. 
It is well known that the pseudogap behavior is pronounced in underdoped regions \cite{IshidaBi2221,Fujiwara}; therefore, the fact that all samples show the reduction of $K_s(T)$ from far above $T_c$ assures that the three samples of 0223F are in underdoped regions. 

In cuprates, the steep decrease of $K_s(T)$ below $T_c$ evidences the reduction in spin susceptibility due to the formation of spin-singlet Cooper pairing; the $T$-derivative of $K_s(T)$, $dK_s(T)/dT$, shows a peak at $T_c$. 
For the present samples, $dK_s(T)/dT$ is shown in Fig.~\ref{fig:KS} as open circles ($\circ$) for OP and as crosses ($+$) for IP. As shown in the figure, $dK_s(T)/dT$ shows two peaks, which correspond to the SC transitions inherent to OP and IP. It has been reported that OP and IP have different $T_c$ values in the same compound due to the charge imbalance between OP and IP, and that the bulk $T_c$ is determined by the higher $T_c$ \cite{Tokunaga}. In Fig.~\ref{fig:KS}, the peak in $dK_s(T)/dT$ at higher temperature is for OP, because $p$(OP) is larger than $p$(IP). 
The SC transition at OP triggers the bulk SC transition; the higher peak coincides with the bulk $T_c$ determined from the onset of SC diamagnetism listed in Table \ref{t:ggg}.

\subsection{Zero-field NMR experiment to evidence an AFM order}

The IP in 0223F($\sharp$3) is the most underdoped CuO$_2$ layer in the present study (see Table~\ref{t:ggg}). The Cu-NMR spectrum of the underdoped IP is lost at low temperatures, as shown in Fig.~\ref{fig:NMR}(c). This is because AFM correlations strongly develop upon cooling towards an AFM order that takes place at low temperatures. NMR measurements at $H_{ex}=0$ T can detect the evidence of an AFM order as explained below.

In general, the Hamiltonian for Cu nuclear spin ($I=3/2$) with an axial symmetry is described in terms of the Zeeman interaction ${\cal H}_{Z}$ due to a magnetic field $H$, and the nuclear-quadrupole interaction ${\cal H}_{Q}$ as follows:
\begin{eqnarray}
{\cal H}&=&{\cal H}_Z+{\cal H}_Q  \notag \\
        &=&-\gamma_N \hbar {\bm I} \cdot {\bm H}+\frac{e^{2}qQ}{4I(2I-1)}(3I_{z^{\prime}}^2-I(I+1)),
\label{eq:hamiltonian}
\end{eqnarray}
where $eQ$ is the nuclear quadrupole moment, and $eq$ is the electric field gradient at a Cu site. In ${\cal H}_{Q}$, the nuclear quadrupole resonance (NQR) frequency is defined as $\nu_{Q}=e^{2}qQ/2h$.
In non-magnetic substances, an NQR spectrum is observed due to the second term in Eq.~(\ref{eq:hamiltonian}) when $H$=$H_{ex}$=0 T. On the other hand, in magnetically ordered substances, an internal magnetic field $H_{int}$ is induced at Cu sites; in addition to the second term, the first term in Eq.~(\ref{eq:hamiltonian}) contributes to the nuclear Hamiltonian even if $H_{ex}$=0 T. Therefore, the onset of a magnetically ordered state is identified from a distinct change of the spectral shape at $H_{ex}$=0 T.

\begin{figure}[htpb]
\begin{center}
\includegraphics[width=0.85\linewidth]{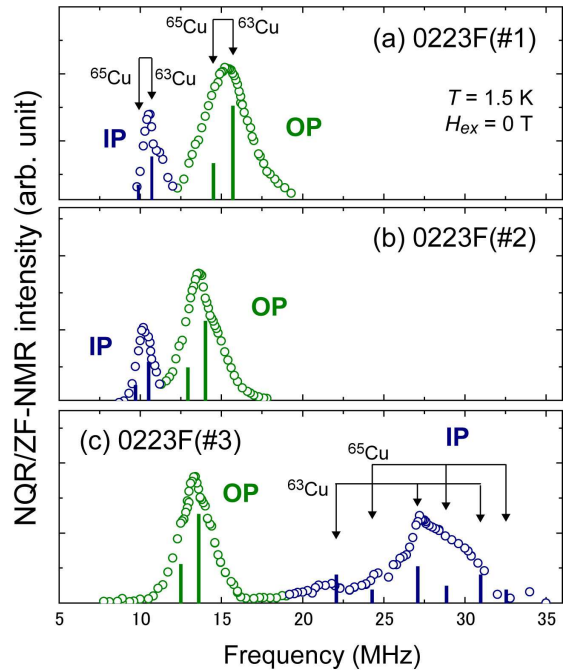}
\end{center}
\caption{\footnotesize (color online)  Cu-NQR or zero-field NMR spectra at $T$=1.5 K for (a) 0223F($\sharp$1) (cited from Ref.~\cite{ShimizuFIN}), (b) 0223F($\sharp$2), and (c) 0223($\sharp$3). The values of $^{63}\nu_Q$ decrease with the reduction of $p$. The solid bars indicate two components of the isotopes, $^{63}$Cu and $^{65}$Cu.}
\label{fig:NQR}
\end{figure}

Figure \ref{fig:NQR}(a) presents the Cu-NQR spectrum of 0223F($\sharp$1), which is observed at $H_{ex}$=0 T. The two peaks correspond to OP and IP, and each peak includes two NQR spectra of $^{63}$Cu and $^{65}$Cu.  
The NQR frequencies are $^{63}\nu_Q$(IP)=10.7 MHz at IP and $^{63}\nu_Q$(OP)=15.5 MHz at OP, which are comparable with the values reported for other multilayered cuprates in the previous literature \cite{ZhengTl2223,ShimizuJPSJ,Julien,Tokunaga,Kotegawa2004}. The observation of the NQR spectrum at $H_{ex}$=0 T assures that 0223F($\sharp$1) is a paramagnetic superconductor. 
Figure~\ref{fig:NQR}(b) shows the Cu-NQR spectrum for 0223F($\sharp$2). The NQR frequencies $^{63}\nu_Q$(IP)=10.4 MHz and $^{63}\nu_Q$(OP)=14.0 MHz are smaller than those for 0223F($\sharp$1). 
This reduction of $^{63}\nu_Q$ comes from the reduction of $p$ in 0223F($\sharp$2) as listed in Table \ref{t:ggg}. In hole-doped cuprates, $^{63}\nu_Q$ decreases with the reduction of $p$~\cite{Ohsugi,Pennington,Yoshinari,ShimizuP,Zheng,Haase}. 

Figure~\ref{fig:NQR}(c) shows the spectrum at $H_{ex}$=0 T for 0223F($\sharp$3). The NQR spectrum is observed for OP, and the NQR frequency $^{63}\nu_Q$(OP)=13.6 MHz is smaller
than those for OP in 0223F($\sharp$1) and 0223F($\sharp$2) due to the reduction of $p$. 
On the other hand, the spectrum of IP is totally different from the NQR spectra for IP in 0223F($\sharp$1) and 0223F($\sharp$2); the resonance frequency of IP in 0223F($\sharp$3) is significantly larger than the NQR frequency expected for the IP, $\sim$ 10 MHz. 
According to Eq.~(\ref{eq:hamiltonian}), resonance frequencies increase when $H_{int}$ is induced at Cu sites by an onset of an AFM order. 
The analysis of the IP's spectrum, assuming Eq.~(\ref{eq:hamiltonian}) and $H_{int} \perp c$, reveals that $H_{int}$ $\sim$ 2.4 T is induced by spontaneous AFM moments $M_{AFM}$ due to the AFM order at IP, as displayed by the bars in the figure.
The obtained value of $H_{int}$ $\sim$ 2.4 T allows us to estimate $M_{AFM}$ $\sim$ 0.12 $\mu_{B}$. Here, we use the relation of $H_{int}=|A_{hf}|M_{AFM}=|A-4B|M_{AFM}$, where $A$ and $B$ are the on-site hyperfine field and the supertransferred hyperfine field, respectively. $A$ $\sim$ 3.7 and $B$ $\sim$ 6.1 T/$\mu_{B}$ are assumed, which are typical values in multilayered cuprates in underdoped regions \cite{ShimizuP}. 

\subsection{Spin-gap behaviors in 1/$T_1T$ at $^{63}$Cu sites}

\begin{figure}[htpb]
\begin{center}
\includegraphics[width=1.0\linewidth]{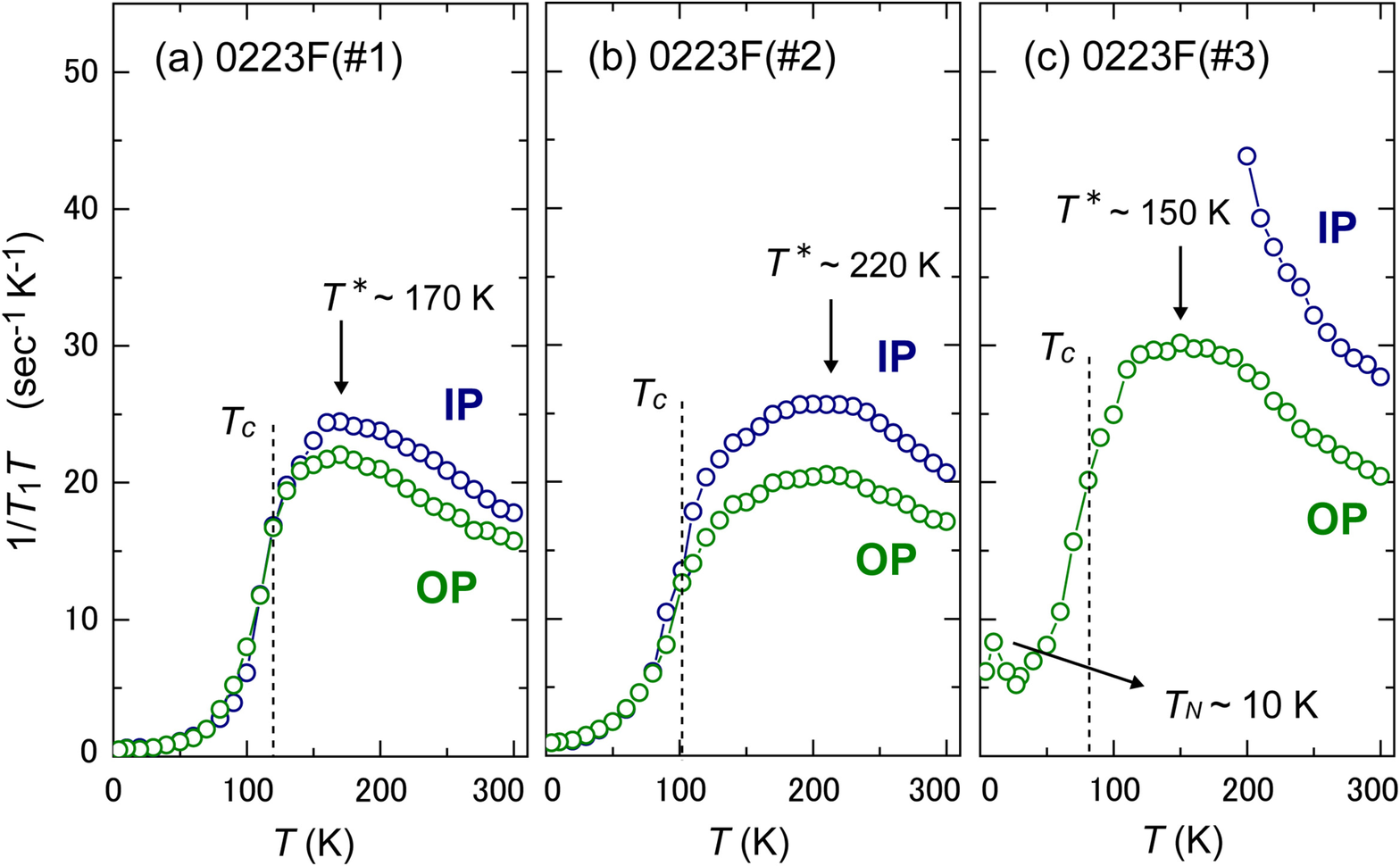}
\end{center}
\caption{\footnotesize (color online)  $T$-dependences of 1/$T_1T$ for (a) 0223F($\sharp$1), (b) 0223F($\sharp$2), and (c) 0223($\sharp$3). The $T_1$ of $^{63}$Cu for OP and IP was measured at $H_{ex}\perp c$ and $\omega_0$=174.2 MHz.}
\label{fig:1/T1T}
\end{figure}
\begin{figure}[htpb]
\begin{center}
\includegraphics[width=1.0\linewidth]{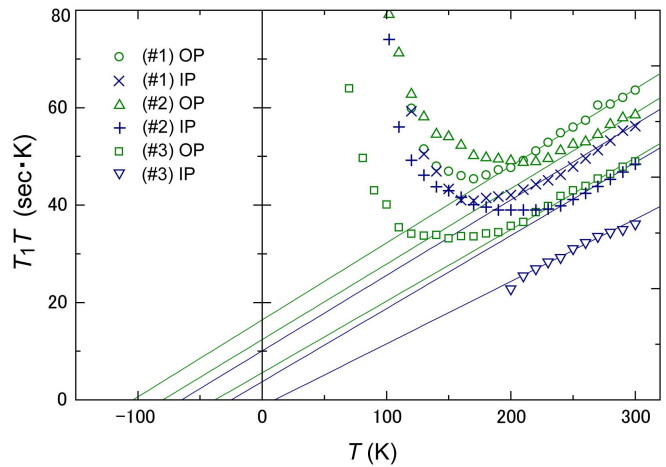}
\end{center}
\caption{\footnotesize (color online) $T$-dependence of $T_1T$ in 0223F. The solid lines represent a Curie Weiss law, $T_1T\sim$ ($T$+$\theta$), where $\theta$ is obtained from the extrapolation of the solid lines at $T$=0. Note that the temperatures which cut the x-axis are $-$$\theta$. The value of $\theta$ decreases with decreasing $p$ and changes its sign to $\theta\sim$ $-$10 K in the IP of 0223F($\sharp$3). This suggests the onset of the AFM order in the IP of 0223F($\sharp$3), which is collaborated with the peak in 1/$T_1T$ for the OP of 0223F($\sharp$3) (see Fig.~\ref{fig:1/T1T}(c)). }
\label{fig:T1T}
\end{figure}

 Figures \ref{fig:1/T1T}(a), \ref{fig:1/T1T}(b), and \ref{fig:1/T1T}(c) show the $T$-dependences of 1/$T_1T$ for 0223F($\sharp$1), 0223F($\sharp$2), and 0223F($\sharp$3), respectively. The nuclear spin-lattice relaxation time $T_1$ was measured at $\omega_0$=174.2 MHz and with $H_{ex}\perp~c$ axis. The quantity 1/$T_1T$ is generally expressed as
\begin{equation}
\frac{1}{T_1T} = \frac{\gamma_N^2 k_B}{2 \mu_B^2} \sum_q  A_{\bm q}A_{-{\bm q}}\frac{{\rm Im}[\chi_{\perp}({\bm q},\omega_0)]}{\omega_0}, 
\label{eq:T1}
\end{equation}
where $A_{\bm q}$ is the ${\bm q}$-component of hyperfine field $A_{hf}$, and $\chi_{\perp}({\bm q},\omega)$ is the perpendicular component of dynamical susceptibility~\cite{Moriya1}.
As shown in the figures, 1/$T_1T$ for IP is larger than that for OP at high temperatures.  According to Eq.~(\ref{eq:T1}), this enhancement of 1/$T_1T$ for IP suggests that low-energy components in $\chi_{\perp}({\bm q},\omega)$ develop with decreasing $p$ as in the case of other hole-doped cuprates \cite{Ohsugi,IshidaBi2221,Fujiwara,Julien,Magishi,ZhengTl2223,Kotegawa2002}.
In 0223F($\sharp$1) and 0223F($\sharp$2), the 1/$T_1T$ exhibits a broad maximum for both IP and OP upon cooling, decreasing from above $T_c$.
By contrast, in 0223F($\sharp$3), the 1/$T_1T$ for IP continues to increase with decreasing $T$. Below $T\sim$ 200 K, it is impossible to measure $T_1$ because the NMR spectrum is lost due to an extremely short relaxation time, as shown in Fig.~\ref{fig:NMR}(c). 
As we have presented in the previous section, the IP in 0223F($\sharp$3) is in an AFM state at $T$=1.5 K; therefore, the continuous enhancement of 1/$T_1T$ is associated with the onset of the AFM order at low temperatures. When a system is in close proximity to the quantum critical point $p_c$ for an AFM order, a two-dimensional AFM spin fluctuation model predicts an expression of 1/$T_1T$ as follows:
\begin{eqnarray}
\frac{1}{T_1T} & \sim & (A-4B)^2\frac{{\rm Im}[\chi_{\perp}({\bm Q},\omega_0)]}{\omega_0} \notag \\
               & \sim & \frac{1}{T+\theta},
\label{eq:T1_2}
\end{eqnarray}
where $\theta$ represents the proximity of a system to $p_c$ and becomes zero at $p_c$ \cite{Millis,Moriya2,Moriya3}.
Thus, the Curie-Weiss law of 1/$T_1T$ is expected. Figure~\ref{fig:T1T} shows the plot of $T_1T$ as a function of $T$, revealing that the value of $\theta$ decreases with $p$. Note that $\theta\sim$ $-$10 K for the IP in 0223F($\sharp$3) points to the onset of the AFM order at $T_N\sim 10$ K, which is also corroborated by the fact that 1/$T_1T$ for the OP in 0223F($\sharp$3) exhibits a peak at $T\sim$ 10 K, as shown in Fig.~\ref{fig:1/T1T}(c). These observations indirectly probe the onset of the AFM order at the IP in 0223F($\sharp$3).

The reduction in 1/$T_1T$ above $T_c$, which is known as a spin-gap~\cite{Yasuoka,REbook}, suggests a gap-opening in low-energy magnetic excitations below the pseudogap temperature $T^*$. In single- and bi-layered compounds, $T^{*}$ increases with decreasing $p$ \cite{IshidaBi2221,ZhengPG}.
On the other hand, in three-layered 0223F, $T^{*}$ for OP coincides with that for IP, despite the charge imbalance between OP and IP, i.e., $p$(OP)$>p$(IP). 
Similar behaviors are reported in other three-layered high-$T_c$ cuprates such as Hg1223~\cite{Julien,Magishi}, Tl1223 \cite{Kotegawa2002}, and Tl2223 \cite{ZhengTl2223}, where OP and IP show the spin-gap in 1/$T_1T$ at the same $T$. 
Especially, optimally-doped \cite{Magishi} and underdoped Hg1223 \cite{Julien} show almost the same $T$-dependence of 1/$T_1T$ with 0223F($\sharp$1) and 0223F($\sharp$2), respectively. 
It remains unresolved at present that $T^{*}$ for IP coincides with that for OP.

As in the case of single- and bi-layered compounds, $T^{*}$ increases with decreasing $p$ from 0223F($\sharp$1) in Fig.~\ref{fig:1/T1T}(a) to 0223F($\sharp$2) in Fig.~\ref{fig:1/T1T}(b). 
In the most underdoped 0223F($\sharp$3), the $T$-dependences of 1/$T_1T$ for OP and IP show unexpected behaviors, although it is expected that those have a peak above the $T^{*}$ $\sim$ 220 K of 0223F($\sharp$2).
As for the IP, 1/$T_1T$ continues to increase upon cooling as shown in Fig.~\ref{fig:1/T1T}(c), suggesting no gap-opening in low-energy components in $\chi_{\perp}({\bm q},\omega)$. The reason for the gapless behavior for the IP of 0223F($\sharp$3) is probably related to the fact that the AFM order takes place at $T_N\sim$ 10 K; it is suggested that a spin-gap diminishes in the underdoped region where an AFM order emerges as the ground state. 
As for the paramagnetic OP, 1/$T_1T$ exhibits a spin-gap behavior below $T^{*}\sim$ 150 K, as shown in Fig.~\ref{fig:1/T1T}(c); however, the value is lower than the $T^{*}$ $\sim$ 220 K of 0223F($\sharp$2), despite the fact that $p$(OP) in 0223F($\sharp$3) is comparable to $p$(IP) in 0223F($\sharp$2), as listed in Table.~\ref{t:ggg}. 

In NMR studies on cuprates, it is a practice to define $T^{*}$ as the $T$ at which 1/$T_1T$ reaches a maximum and identify it as the $T$ below which the low-energy spectral weight in magnetic excitations is suppressed. Instead, as the onset $T$ of the spin-gap behavior, it is also possible to assign $T^{*}$(CW); it is defined as the $T$ at which the $T$ dependence of $T_1T$ deviates from the Curie-Weiss law. 
As shown in Fig.~\ref{fig:T1T}, $T^{*}$(CW) $\sim$ 210 K for the OP in 0223F($\sharp$3) is relatively close to $T^{*}$(CW) $\sim$ 240 K for the IP in 0223F($\sharp$2), which is consistent with the fact that the $p$ values are comparable.
Such instabilities of the spin-gap temperature imply that the peak in 1/$T_1T$ results from the competing effects; AFM spin fluctuations increasing 1/$T_1T$ and spin-gap behaviors suppressing 1/$T_1T$. The peak in 1/$T_1T$ would then occur at the $T$ at which the spin-gap behavior begins to win the competition. 
In this context, as a possible explanation of the low $T^{*}$ value for the OP in 0223F($\sharp$3), we may consider that the real $T^{*}$ value for the OP is masked by the huge enhancement of 1/$T_1T$ for the IP. There is the supertransferred hyperfine coupling between Cu nuclei at OP and Cu-derived spins at IP, which increases 1/$T_1T$ at the OP and hence decreases $T^{*}$. As a result, it is likely that $T^{*}$ at the OP is seemingly decreased down to around 150 K.

In another case, we should note that a reduction of $T^{*}$ has been also reported in underdoped YBCO near the boundary of a magnetic phase \cite{Berthier}.
This suggests that $T^{*}$ is reduced as a system approaches a magnetic phase boundary, even when there is no adjacent IP that shows an AFM ordering. In any case, further works are required to investigate magnetic excitations in the vicinity of $p_c$ where an AFM order dose not set in at low temperatures.


\section{Discussions}

\subsection{AFM and SC phase diagram of three-layered 0223F}
\begin{figure}[htpb]
\begin{center}
\includegraphics[width=1.0\linewidth]{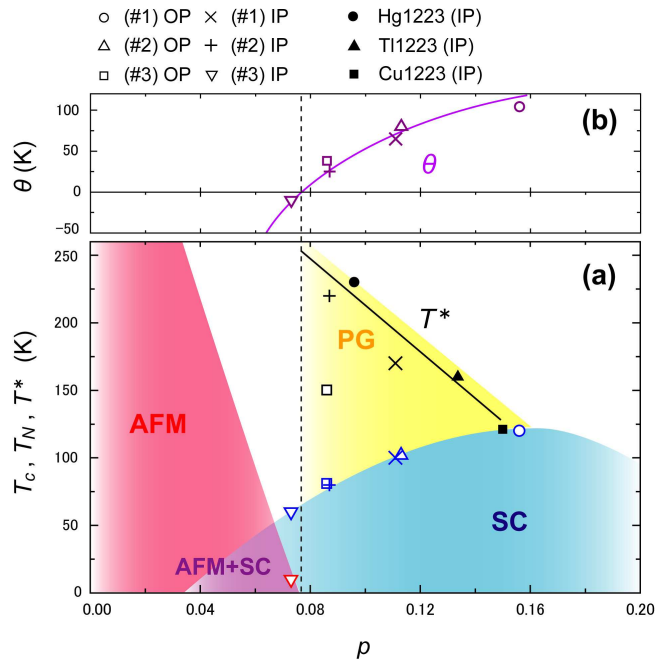}
\end{center}
\caption{\footnotesize (color online)  (a) Phase diagram of three-layered Ba$_2$Ca$_2$Cu$_3$O$_6$(F,O)$_2$. $T_c$ and $p$ were determined by the Knight Shift; $T_N$ and $T^{*}$ were determined by 1/$T_1T$. In the figure, the data for OP and IP are separately plotted. The data of $T^{*}$ include other three-layered compounds: Hg1223 \cite{Julien}, Tl1223 \cite{Kotegawa2002}, and Cu1223 \cite{unpublished}. PG denotes the pseudogap phase. (b) We show the $p$-dependence of $\theta$, the values of which were obtained from Fig.~\ref{fig:T1T}. The quantum critical point $p_c$ for the AFM order, $p_c$ $\simeq$ 0.075, is determined as the hole density at which $\theta$ is zero.}
\label{fig:PD}
\end{figure}
Figure \ref{fig:PD}(a) shows the phase diagram of three-layered 0223F. The data for OP and IP are separately plotted in the figure. 
With decreasing $p$, $T_c$ gradually decreases, and the AFM phase appears at $p$ $\sim$0.07-0.08. 
Here, we determine the quantum critical point $p_c$ for the AFM order, $p_c$ $\sim$ 0.075, as the hole density at which $\theta$ is zero (see Fig.~\ref{fig:PD}(b)).
This phase diagram suggests that three-layered 0223F has a coexistence phase of AFM and SC
in an underdoped region, as reported in four-\cite{ShimizuJPSJ} and five-layered compounds \cite{MukudaJPSJ,ShimizuFIN}.
Note that the coexistence of AFM and SC is not a phase separation between magnetic and paramagnetic phases; a long range AFM order and SC uniformly coexist in a CuO$_2$ plane \cite{ShimizuFIN}.
In the case that such a phase separation occurs, a paramagnetic NQR spectrum should be observed at around 10 MHz in the zero-field NMR spectra shown in Fig.~\ref{fig:NQR}(c). 
The critical point in 0223F, $p_c$ $\simeq$ 0.075, is larger than those in LSCO, $p_c$ $\simeq$ 0.02 \cite{Keimer,JulienLSCO}, and in YBCO, $p_c$ $\simeq$ 0.055 \cite{Sanna,Coneri}; $p_c$ increases with the stacking number $n$ of CuO$_2$ layers in a unit cell. 
This is qualitatively because the magnetic interlayer coupling becomes stronger by increasing $n$. When we increase $n$ more than three, $p_c$ becomes larger and reaches $p_c$ $\sim$ 0.08-0.09 in four-layered compounds \cite{ShimizuJPSJ,p} and $p_c$ $\sim$ 0.10-0.11 in five-layered ones \cite{MukudaJPSJ,p}.
These results suggest that the uniform coexistence of AFM and SC is a universal phenomenon in underdoped regions, when a magnetic interlayer coupling is strong enough to stabilize an AFM order. As for these experimental results, a recent theoretical study seems to support this conclusion \cite{Yamase}.

\subsection{Pseudogap in three-layered 0223F}

Since the discovery of high-$T_c$ cuprates, anomalous metallic states with pseudogap behaviors have been one of the most exciting and important subjects in condensed-matter physics. 
Initially, a gap-like behavior was reported as a gradual suppression of $1/T_1T$ below $T^{*}$ \cite{Yasuoka,Warren,Horvatic}, which has been called {\it spin-gap}. Neutron scattering experiments in underdoped regions also showed that spin excitations at low energies were suppressed in the normal state~\cite{Rossat}. Then, ARPES experiments directly identified the existence of the energy gap in electronic spectra even above $T_c$~\cite{Loeser,Ding}. This gap, observed in ARPES below $T^{*}$, turned out to have the same angular dependence as the $d$-wave SC gap in the Brillouin zone \cite{Ding,Harris}. After these experimental observations on single particle spectra, the gap has been called {\it pseudogap}.   
  
\begin{figure}[htpb]
\begin{center}
\includegraphics[width=1\linewidth]{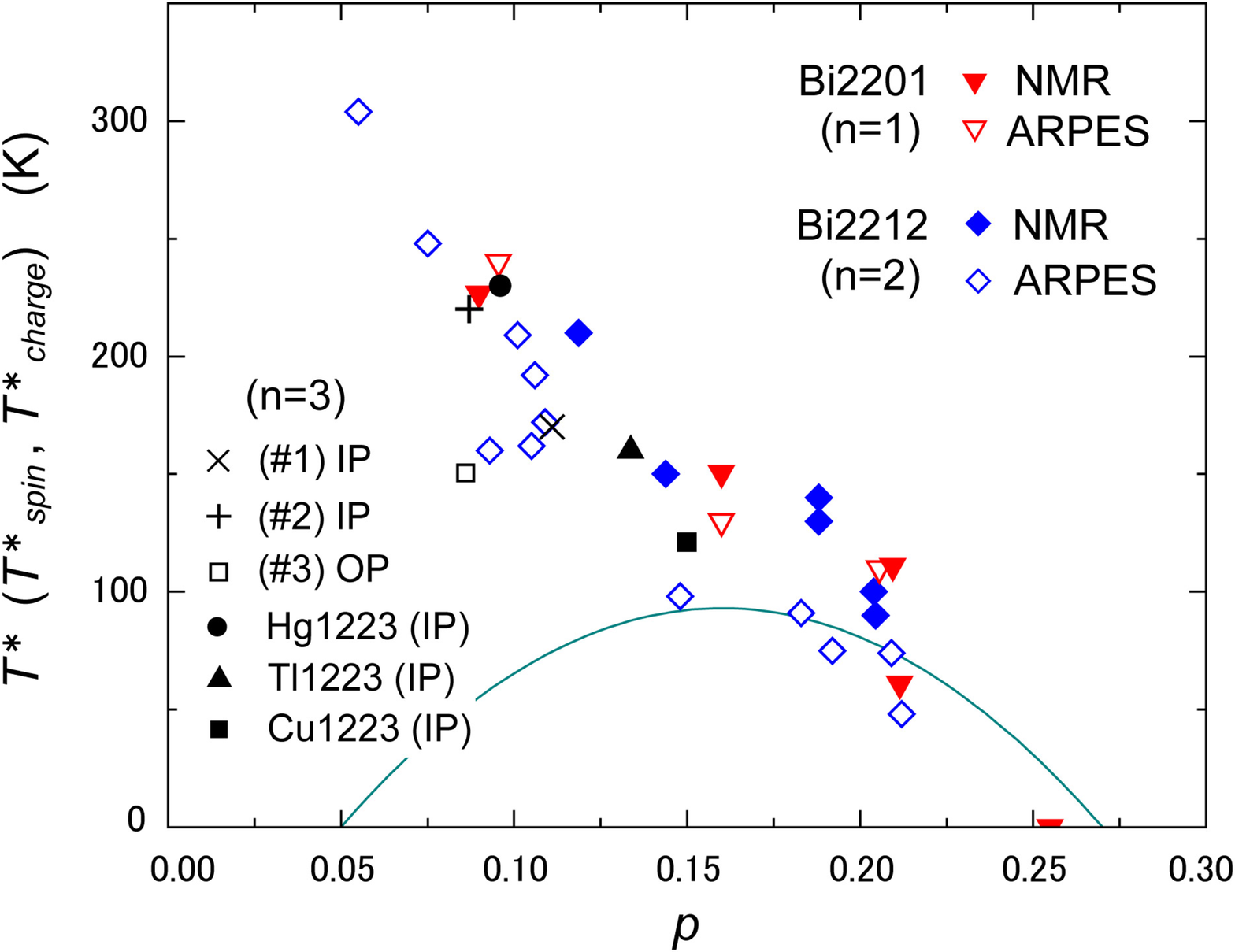}
\end{center}
\caption{\footnotesize (color online)  $p$-dependences of $T^{*}$ for single-layered Bi2201 ($n$=1) and bi-layered Bi2212 ($n$=2). $T^{*}$ were determined with ARPES~\cite{Kondo,Campzano} and NMR~\cite{ZhengPG,IshidaBi2221,Walstedt}. The solid line represents the SC phase for Bi2212 as $T_c$=$T_{c,max}$[1-82.6(1-0.16)$^2$] \cite{Groen,Presland}. The data for three-layered compounds ($n$=3) are the same as those in Fig.~\ref{fig:PD}(b).}.
\label{fig:PG}
\end{figure}

Figure \ref{fig:PG} shows the $p$-dependence of $T^{*}$ for single-layered Bi2201 ($n$=1) \cite{Kondo,ZhengPG} and bi-layered Bi2212 ($n$=2) \cite{Campzano,IshidaBi2221,Walstedt}, the values of which were reported by NMR and ARPES studies. 
Henceforth, we denote $T^{*}$ determined by NMR and ARPES as $T^{*}_{spin}$ and $T^{*}_{charge}$, respectively. 
Note that in Fig.~\ref{fig:PG} both $T^{*}_{spin}$ and $T^{*}_{charge}$ remain unchanged between $n$=1 and $n$=2, as mentioned in literature \cite{Yoshida}. Moreover, it is noteworthy that $T^{*}_{spin}$ coincides with $T^{*}_{charge}$, which suggests that the magnetic and charge excitations are suppressed below the same $T$.  
As for the three-layered compounds ($n$=3), the $T^{*}_{spin}$ values of 0223F are presented in Figs.~\ref{fig:PD}(a) and \ref{fig:PG}, along with the data of other three-layered compounds such as Hg1223~\cite{Julien}, Tl1223~\cite{Kotegawa2002}, and Cu1223~\cite{unpublished}. Here, except for 0223F($\sharp$3), we plot the $T^{*}_{spin}$ values only for IP.
As shown in Fig.~\ref{fig:PG}, the $p$ dependence of $T^{*}_{spin}$ for $n$=3 coincides with $T^{*}_{spin}$ and $T^{*}_{charge}$ for $n$=1 and 2.
Despite the stronger magnetic interlayer coupling in $n$=3 than in $n$=1 and 2, the $p$-dependence of $T^{*}$ is unchanged from those in $n$=1 and 2; it seems that $T^{*}$ is determined by {\it in-plane} magnetic and charge correlations \cite{Yoshida}. 
From a theoretical point of view, the $t-J$ model has explained $T^{*}$ as an onset temperature for the singlet pairing of spinons~\cite{Ogata}, which would be insensitive to the presence of the magnetic interlayer coupling. 
Here, as discussed before, note that the plot for the OP in 0223F($\sharp$3) is inconsistent with other data as shown in Figs.~\ref{fig:PD}(a) and \ref{fig:PG}; detail discussions on this matter remains as a future work.

Finally, we highlight the fact that the spin-gap behavior disappears when $p < p_c$; 1/$T_1T$ for the IP in 0223F($\sharp$3) continues to increase upon cooling, as shown in Fig.~\ref{fig:1/T1T}(c).  
In underdoped regions where the ground state is the coexistence of SC and AFM, the low-lying magnetic excitations develop toward the AFM order upon cooling, without the gap-opening related to the development of spin-singlet formation above $T_c$. 
Recently, Yamase {\it et al.} have theoretically pointed out that a spin-gap is strongly suppressed near a tetra-critical point between AFM, SC, AFM+SC, and a normal state~\cite{Yamase}. 
In this context, it remains possible that a pseudogap, which emerges around the antinode region at the wave vectors (0,$\pi$) and ($\pi$,0) when $p$ $>$ $p_c$, evolves to a real {\it AFM gap} when $p$ $<$ $p_c$. 
The underlying issue is to experimentally address whether $T^{*}_{charge}$ survives or not in the paramagnetic normal state in the AFM-SC mixed region where $T^{*}_{spin}$ disappears. 

\section{Conclusion}

Cu-NMR studies on three-layered Ba$_2$Ca$_2$Cu$_3$O$_6$(F,O)$_2$ have unraveled the phase diagram of AFM and SC, where AFM uniformly coexists with SC in an underdoped region.
The critical point $p_c$ for the AFM order is determined to be $p_c$ $\simeq$ 0.075 in
the three-layered compounds; the value is larger than in single- and bi-layered compounds, and smaller than in four- and five-layered ones.
This is because the magnetic interlayer coupling becomes stronger as the stacking number $n$ of CuO$_2$ layers increases.
These results indicate that the uniform coexistence of AFM and SC is a universal phenomenon in an underdoped region when a magnetic interlayer coupling is strong enough to stabilize an AFM
ordering.
We have highlighted the fact that the $p$ dependence of the spin-gap temperature $T^{*}$ is independent of $n$ when $p$ $>$ $p_c$, and that the spin-gap collapses in the AFM region, i.e., $p < p_c$.
The unexpected reduction of $T^{*}$ for the OP in 0223F($\sharp$3) remains to be investigated.

\section*{Acknowledgement}

The authors are grateful to M. Mori for his helpful discussions. This work was supported by Grant-in-Aid for Specially promoted Research (20001004) and for Young Scientists (B) (23740268) and  by the Global COE Program (Core Research and Engineering of Advanced Materials-Interdisciplinary Education Center for Materials Science) from The Ministry of Education, Culture, Sports, Science and Technology, Japan.


\clearpage

\end{document}